\documentclass[pra,twocolumn,showpacs,amsmath,amssymb,floatfix]{revtex4}

\usepackage{graphicx}

\begin{document}
\title{Study of the Born-Oppenheimer Approximation \\
for Mass-Scaling of Cold Collision Properties}

\author{Stephan Falke}
\author{Eberhard Tiemann}
\author{Christian Lisdat}
 \email{lisdat@iqo.uni-hannover.de}

\affiliation{Institut f{\"u}r Quantenoptik, Leibniz Universit{\"a}t Hannover; Welfengarten 1; 30167 Hannover; Germany}

\date{\today}

\begin{abstract}
Asymptotic levels of the A~$^1\Sigma_u^+$ state of the two isotopomers $^{39}{\rm K}_2$ and $^{39}{\rm K}^{\,41}{\rm K}$ up to the dissociation limit are investigated with a Doppler-free high resolution laser-spectroscopic experiment in a molecular beam. The observed level structure can be reproduced correctly only if a mass dependent correction term is introduced for the interaction potential. The applied relative correction in the depth of the potential is $10^{-6}$, which is in the order of magnitude expected for corrections of the Born-Oppenheimer approximation. A similar change in ground state potentials might lead to significant changes of mass-scaled properties describing cold collisions like the s-wave scattering length.
\end{abstract}

\maketitle

\section{Introduction}
\label{intro}
The field of experiments with ultracold neutral atoms is evolving fast. Tests of fundamental physics have been performed and many more are suggested. Examples are the realization of a Bose-Einstein condensate (BEC) \cite{anderson95,davisa95}, a quantum degenerate fermi gas \cite{demarco99,modugno02}, or a Mott-insulator state \cite{greiner02}. It is essential to understand and manipulate cold collisions, e.g., by magnetic fields, which allow to set the effective interaction strength at will.

The effective two-body interaction strength at the threshold is usually described by the s-wave scattering length $a_{\rm s}$ \cite{roberts01}. It can be altered from an effectively attractive ($a_{\rm s}<0$) one to an effectively repulsive one ($a_{\rm s}>0$) in the vicinity of Feshbach resonances. Here, scattering states that describe two colliding atoms of an ultracold ensemble are coupled typically by hyperfine interaction to bound levels of the corresponding molecule. An example for the utilization of the manipulation of the interaction strength is the investigation of a BEC-BCS crossover \cite{bourdel04,greiner05}.

The scattering length of a binary collision corresponds to a phase shift that the scattering wavefunction experiences due to the influence of the potential describing the interaction of one atom colliding with the other. Therefore, the scattering length or phase shift reduces the effect of the interaction potential to a single parameter. A way to obtaining precise values of the scattering length and Feshbach resonance positions is to derive a full interaction potential curve~$V\left(R\right)$ from molecular spectroscopy ($R$ being the internuclear separation) \cite{samuelis01}. It is also possible to derive quantitative models for cold collisions from spectroscopy of the last bound levels by asymptotic methods \cite{verhaar93,crubellier99}.

The interaction potential curve of two neutral atoms is due to the electronic interactions of electrons and the two nuclei. Changing the number of neutrons in the nuclei should therefore not alter the interaction potential curve if the nuclei are regarded as infinitely heavy in comparison with the electrons. This is expressed by the Born-Oppenheimer approximation. However, a change in the reduced mass of the system of two atoms will alter the vibrational motion in the potential and rotational level structure due to the modified centrifugal barrier. Thus, the position of the bound levels of the potential becomes mass dependend resulting in variations of the scattering length from one isotope to another by changes of the accumulated scattering phase. But these changes are obtained in a straight forward way by solving the Schr{\"o}dinger equation of the nuclear motion with the appropriate reduced mass.

The effective number of vibrational levels of a potential $V\left(R\right)<0$ with $V\rightarrow 0$ for large $R$ and $V\left(R_{\rm i}\right) = 0$ is related to the total phase of the potential. It is in WKB approximation
\begin{equation}
v_{\rm D}=\sqrt{\mu}\ \frac{\sqrt{8}}{h}\int_{R_{\rm i}}^\infty \sqrt{-V\left(R\right)}\ {\rm d}R
\end{equation}
and scales with the square-root of the reduced mass~$\mu$ of the isotope combination. The non-integer modulo of this quantity is a measure for the phase shift during the collision at threshold and thus relates to the scattering length \cite{gribakin93}. Provided that the number of supported vibrational levels of the potential is known, it is possible to predict the scattering length from one isotope combination to another, a method often referred to as mass-scaling.

But, this method produces only reliable results if the interaction potential curves are mass independent, i.e., the Born-Oppen\-heimer approximation is precisely fulfilled. In molecular spectroscopy the mass-scaling method is regularly used to test the vibrational assignment if term values are available for two isotopomers. It is known that the mass relations and in this sense also the Born-Oppen\-heimer approximation are good to a relative precision of $10^{-5}$ for deeply bound molecular levels \cite{pashov05,stwalley75}. Applying high precision microwave spectroscopy deviations from the Born-Oppen\-hei\-mer approximations were observed in the rotational structure for several examples \cite{watson80,rosenblum58,tiemann82}. It was also observed in optical spectra for PbO \cite{knoeckel84} and iodine dimer \cite{knoeckel04,salumbides06} that corrections are needed if one compares the molecular structure of two isotopomers with very high precision.

In this work we want to study the precision of the Born-Oppenheimer approximation for asymptotic levels, i.e., of levels close to a dissociation limit. We compare high resolution spectra of two potassium isotopomers ($^{39}{\rm K}_2$ and $^{39}{\rm K}^{\,41}{\rm K}$). The experiment is described in Chapter~\ref{experiment}. The underlying theory is discussed in Chapter~\ref{theory} and a comparison of the corresponding simulations and the experimental observations is undertaken in Chapter~\ref{analysis}. We conclude and give an outlook in Chapter~\ref{conclusion}.

\section{Experiment}
\label{experiment}
In a recent publication we reported on a study of the A~$^1\Sigma_u^+$ state of $^{39}{\rm K}_2$ up to the dissociation limit \cite{falke206}. In this work, we focus our interest on the least bound levels of two isotopomers. We use the same experimental setup as before \cite{lisdat01,lisdat201}.

\begin{figure}
\begin{center}
\resizebox{1.0\columnwidth}{!}{\includegraphics{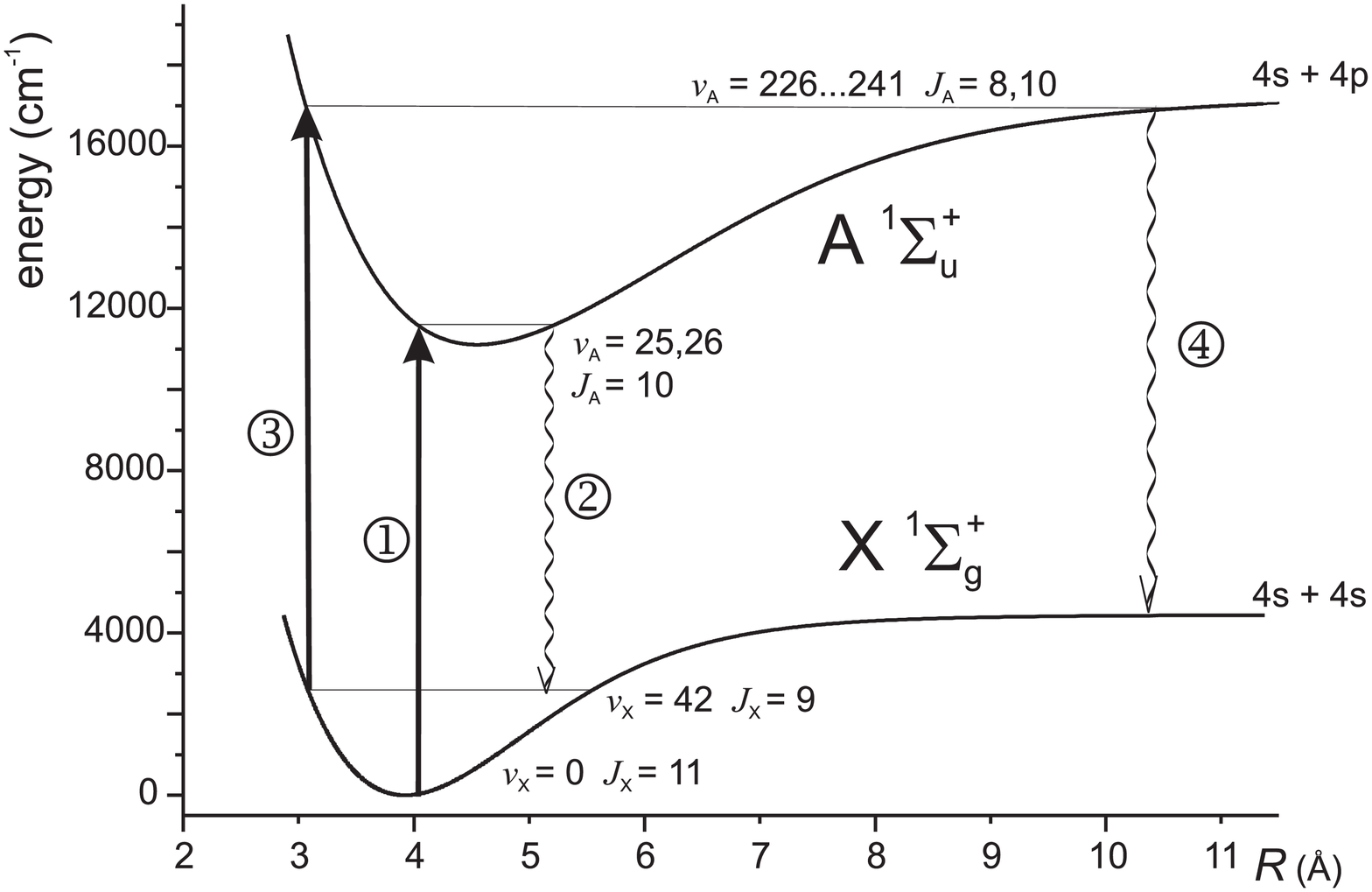}}
\end{center}
\caption{Simplified potential scheme of ${\rm K}_2$. The preparation of ${\rm K}_2$ is done by Franck-Condon pumping  with a fixed frequency laser indicated by \textcircled{1} followed by spontaneous decay indicated by \textcircled{2}. The spectroscopy is done by the scanning laser \textcircled{3}, which induces fluorescence \textcircled{4} mainly to continuum states just above the ground state asymptote. This laser induced fluorescence is the experimental signal.}
\label{figSetup}
\end{figure}

In brief, a highly collimated molecular beam is produced via adiabatic expansion of hot potassium vapor into vacuum. Ground state molecules ($v_{\rm X}=0$, $J_{\rm X}=11$) are optically pumped to a higher vibrational level of the ground state (see Figure \ref{figSetup}, \textcircled{1} and \textcircled{2}): a continuous Ti:sapphire laser drives the P(11) line of the 25--0 band (or 26--0 for $^{39}{\rm K}^{\,41}{\rm K}$) of the A--X system (Franck-Condon pumping).

A second Ti:sapphire laser interacts with the molecules, which decayed to vibrationally excited ground state levels, in a zone spatially separated from the interaction of the preparation laser with the molecules. This spectroscopy laser scans and induces fluorescence from the prepared molecules, which is detected by a photomultiplier (see \textcircled{3} and \textcircled{4} in Figure \ref{figSetup}). A band-pass filter around the potassium $D_1$ line is used in combination with a color-glass filter for the suppression of stray laser light and discrimination of fluorescence from molecular levels, which are not close to the ${\rm s}+{\rm p}$ asymptote.

\begin{figure}
\begin{center}
\begin{tabular}{@{}l@{}}
\resizebox{1.0\columnwidth}{!}{\includegraphics{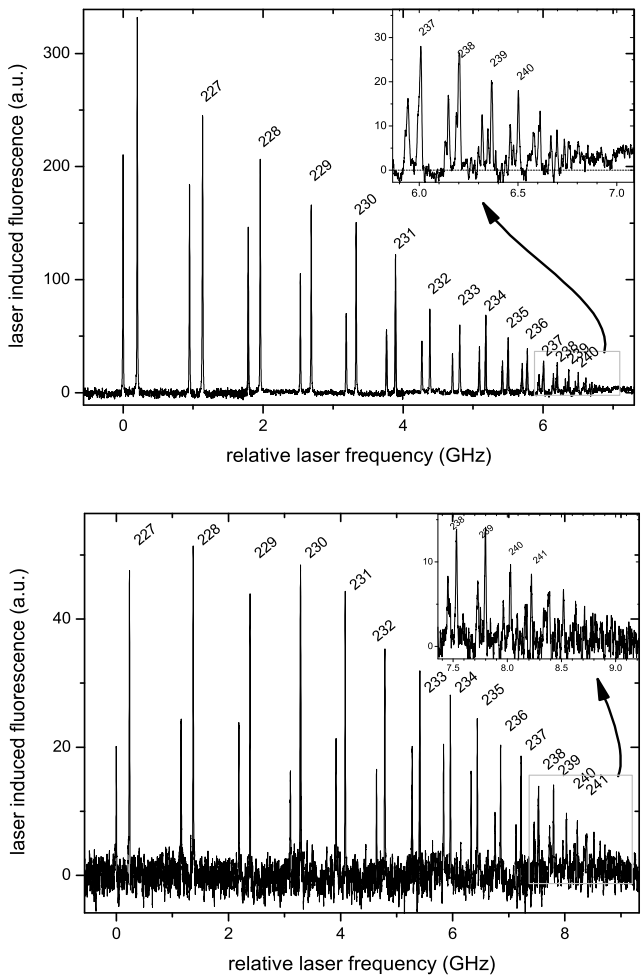}}
\end{tabular}
\end{center}
\caption{Scan across asymptotic levels of the A~$^1\Sigma_u^+$  state of the homonuclear $^{39}{\rm K}_2$ (top graph) and the heteronuclear $^{39}{\rm K}^{\,41}{\rm K}$ dimer (bottom graph). Two rotational lines P(9) and R(9) are visible for each vibrational level (indicated by the number on top of the pair of lines.)}
\label{figScans}
\end{figure}

The absolute frequency of the spectroscopy laser is measured by a wavelength meter with a precision of $100~{\rm MHz}$ (HighFinesse WS/7). The relative frequency axis is obtained with a temperature stabilized Fabry-P{\'e}rot interferometer, which has a free spectral range of $150~{\rm MHz}$. The interpolation between the transmission peaks obtained from this Fabry-P{\'e}rot interferometer is good to about $2~{\rm MHz}$. The precision is limited by non-linearities in the piezoelectric scan of the Ti:sapphire laser. For each scan of the laser we fit a polynomial of forth order to the marker positions to obtain a relative frequency axis (scan range $3~{\rm GHz}$). No systematic drifts of the markers are visible to our precision during the observation time of some hours. In consequence, we average several scans of a frequency interval using the obtained local frequency axis. Also, several frequency intervals can be connected to a single long scan of the frequency of the spectroscopy laser. The result of this procedure is shown in Figure~\ref{figScans}, where a background due to residual stray laser light is removed.

The two scans show laser induced fluorescence from asymptotic levels of the two most abundant isotopomers $^{39}{\rm K}_2$ (87\%) and $^{39}{\rm K}^{\,41}{\rm K}$ (12.5\%). The difference of abundance explains the better signal-to-noise ratio for the ho\-mo\-nuclear $^{39}{\rm K}_2$. Moreover, we used the 26--0 band for Franck-Condon pumping in case of the heteronuclear $^{39}{\rm K}^{\,41}{\rm K}$ instead of the 25--0 band for the homonuclear case because the corresponding band in $^{39}{\rm K}^{\,41}{\rm K}$ is strongly perturbed due to spin-orbit interaction of the A~state with the b~$^3\Pi_u$ state. Therefore, the Franck-Condon factor for the excitation is smaller and the start level was not completely emptied. The scans show rotational doublets for a vibrational progression of the A~state up to the dissociation limit. The energy spacing between vibrational and rotational levels of the ground state is bigger than the interval of observations and all lines start from the same ground state level. A similar progression could be obtained for $J_{\rm X}=11$ with P(11) and R(11) doublets. The highest vibrational levels of the A~state of the homonuclear dimer show hyperfine structure (see Figure~\ref{figScans}, top graph, $v_{\rm A}=238$ and higher). The vibrational numbering is known from our previous study \cite{falke206} exactly for $^{39}{\rm K}_2$. In that study the complete vibrational series was followed from the potential minimum to the asymptote. For the heteronuclear case no hyperfine structure is visible, which is expected from the smaller hyperfine parameters of $^{41}{\rm K}$ as we will discuss below. Also, the onset of the continuum above the asymptote is visible for the homonulcear dimer.

\section{Theory}
\label{theory}
In general, the atomic transition frequencies for two isotopes differ by the isotope shift. Therefore, two asymptotic energies play a role at the, e.g., ${\rm s}+{\rm p}$ asymptote of a dimer constituted from two different isotopes of the same element. If one neglects fine and hyperfine structure, the two atoms interact via an induced dipole-dipole interaction at large internuclear separations, which is described by a $C_6/R^6$ behavior of the interaction potential. In the special case of homonuclear dimers, the two asymptotic energies become degenerate and consequently the interaction is a resonant dipole-dipole interaction, which is described by a $C_3/R^3$ behavior. This special case is well studied by homonuclear alkali metal dimers, in which the atomic radiative lifetime can be calculated from the long range dispersion coefficient~$C_3$ \cite{falke206,tiemann96,wang297}. For internuclear separations where the potential energy becomes larger than the energy difference from the isotope shift the long range behavior goes over to $R^{-3}$ even for heteronuclear dimers if atomic isotope shift is small.

\begin{figure}
\begin{center}
\resizebox{1.0\columnwidth}{!}{\includegraphics{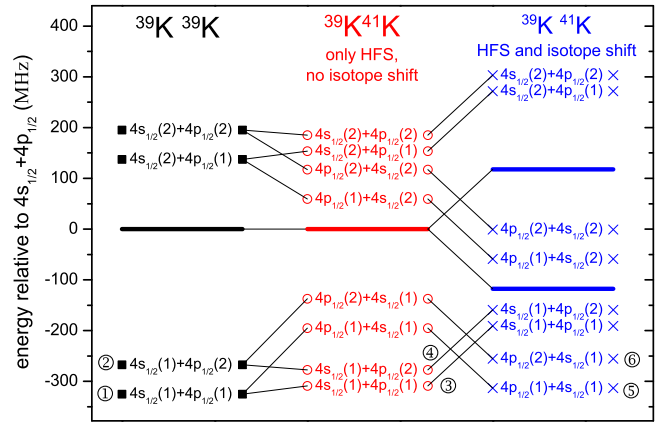}}
\end{center}
\caption{Structure of asymptotic energies of $^{39}{\rm K}_2$ and $^{39}{\rm K}^{\,41}{\rm K}$ on the left and right respectively. The middle shows the artificial situation of the heteronuclear case without isotope shift. The connecting lines indicate the changeover to the homonuclear and the heteronuclear cases. The number in brackets indicates the total angular momentum. The centers of gravity of the hyperfine structure are given by thick lines for the three cases. The two centers of gravity on the right hand side stem from the different excitation energies of $^{39}{\rm K}$ and $^{41}{\rm K}$ due to the isotope shift. The two centers are shifted up and down from the original center by half of the isotope shift of the $D_1$ line, which ensures that the resulting potential curves overlap with those of the two other models at smaller internuclear separations, where the dipole-dipole interaction is the main energy contribution. The circled numbers indicate the asymptotic energies of the A~state.}
\label{figAsympEnergies}
\end{figure}
In this study, the isotope shift between the $D$~lines of the atoms $^{39}{\rm K}$ and $^{41}{\rm K}$ is with $235~{\rm MHz}$ \cite{falke06} in the order of the hyperfine splitting, which itself already perturbs (along with the fine structure) the pure $C_3/R^3$ structure of the homonuclear dimer. In our preceding publication, we discussed how the fine and hyperfine structure of the atoms is taken into account for the derivation of the potential curve \cite{falke206}. The potential energy curve of the A~$^1\Sigma_u^+$ state divides for very large internuclear separations into several curves, each corresponding to a combination of hyperfine states of the atoms. These branches are calculated purely from atomic parameters, namely fine and hyperfine parameters and the $C_3$ parameter.

The same calculations can be done for a heteronuclear dimer by including two different sets of hyperfine parameters and the isotope shift of the transition frequencies \cite{falke06}. Taking the isotope shift into account for the asymptotic behavior already relates to corrections of the Born-Oppenheimer approximation.

In Figure~\ref{figAsympEnergies}, the asymptotic energies of the lower spin-orbit asymptote $4{\rm s}_{1/2}+4{\rm p}_{1/2}$ are shown. On the left hand side, the structure of the homonuclear dimer $^{39}{\rm K}_2$ is shown. In the middle column, the exchange degeneracy is removed for the heteronuclear dimer $^{39}{\rm K}^{\,41}{\rm K}$ because the two atoms have different hyperfine structure. The isotope shift is, however, not considered. The center of gravity of the two $4{\rm p}_{1/2}$ states is at the same energy. The isotope shift (taken from reference \cite{falke06}) is introduced on the right hand side. We distribute the isotope shift on the two atoms: The asymptotic energies with the $^{39}{\rm K}$ atom in the $4{\rm p}$ state are shifted down by half of the isotope shift, the energies with the $^{41}{\rm K}$ atom the $4{\rm p}$ state are shifted upwards by half of the isotope shift. This ensures that the potential energy curves become the same if one goes to smaller internuclear distances. The two energy scales of the two isotopomers are therefore connected by a common energy for the bottom of the A~$^1\Sigma_u^+$ state if no additional mass dependent correction of the Born-Oppenheimer approximation would be needed.

By connecting calculated adiabatic (diagonalizing fine, hyperfine structure, and dipole-dipole interaction) long range potentials to the inner potential obtained in our previous study \cite{falke206}, one is able to calculate the term energies for all levels of the A~state. The atomic hyperfine structure influences the potential curves and is therefore reflected by the molecular hyperfine structure. The observed molecular hyperfine structure is a combined effect of hyperfine splittings of the  atoms in the ground state and in the excited state. Due to the smaller atomic hyperfine structure of $^{41}{\rm K}$ the molecular hyperfine sructure of $^{39}{\rm K}^{\,41}{\rm K}$ is expected to be smaller than for $^{39}{\rm K}_2$. All molecular potential curves of the A~state end at the two lowest asymptotes, which are seperated by the hyperfine splitting of the atomic $4{\rm p}_{1/2}$ state only. Of course, one needs to take the different reduced masses of the dimers into account for the calculation of term energies (see Section~\ref{intro}). In the following, we compare experimental findings and calculations.

\section{Analysis}
\label{analysis}
In the process of the analysis, three steps are taken. They are summarized here:
\begin{enumerate}
\item fine adjustment of the potential of the A~$^1\Sigma_u^+$ state for the asymptotic region of the homonuclear molecule,
\item comparison of the heteronuclear with the homonuclear dimer, and
\item correction of the Born-Oppenheimer potential for the heteronuclear dimer.
\end{enumerate}

\subsection{Homonuclear Dimer}
\begin{figure}
\begin{center}
\resizebox{1.0\columnwidth}{!}{\includegraphics{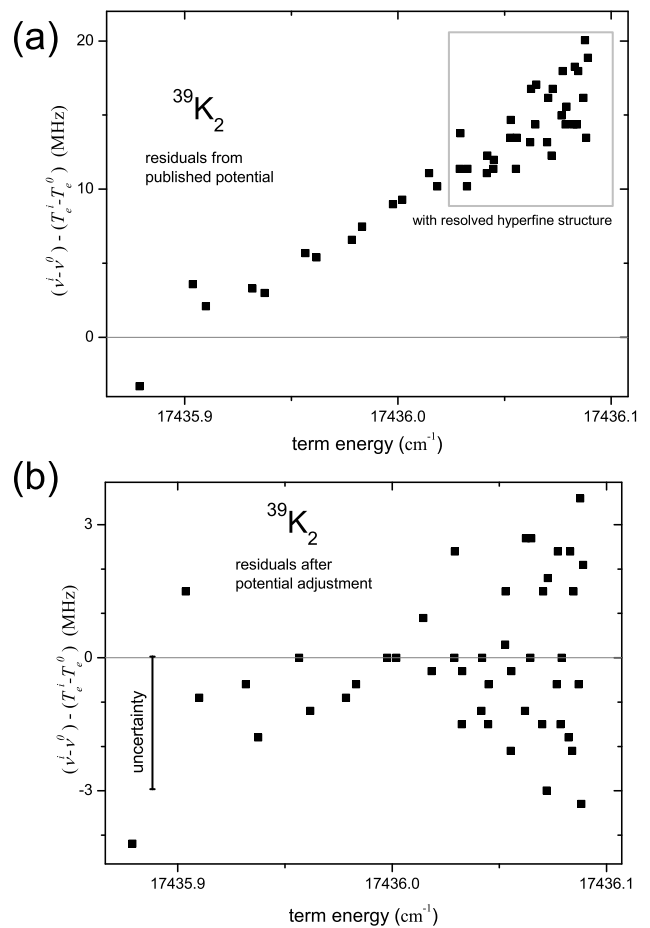}}
\end{center}
\caption{Analysis of the asymptotic term energies of the homonuclear dimer. {\rm Graph~(a)} shows the residuals of the energy spacing resulting from the potential derived in our preceding study \cite{falke206}. After adjusting the inner branch of the potential by a fit, the residuals are reduced as shown in {\rm Graph~(b)}, note the scale change by a factor of about three compared to {\rm Graph~(a)}.}
\label{figResidualHomo}
\end{figure}

For a direct comparison between the two isotopomers, we recorded spectra of the homonuclear molecule starting from the same rotational state, i.e., $J_{\rm X}=9$ (see Figure~\ref{figScans}). The hyperfine structure of the least bound levels of the homonuclear dimer is nicely resolved and can easily be seen, e.g., for $v_{\rm A}\!=\!239$. We assigned the observed lines and determined energy spacings with the FPI marker trace. The absolute frequency measurement is used for the assignment of the reference line $\nu^0$ (226--43~P(9) at $14\,145.105(3)\;{\rm cm}^{-1}$) but not for the analysis itself. The assignment of higher lines ($\nu^i$, $i=1,\dots,N$) is done by simply counting the doublets from the already identified line.

We compared the experimental observations with calculations from the potential derived in our preceding study \cite{falke206}. To do so, a list of measured transition energies $\nu^i$ and a list of calculated term energies $T_e^i$ ($i=0,\dots N$) are derived. From these lists, energy spacings are calculated. The adiabatic picture for the potential curves is still justified since no avoided crossings of the hyperfine potentials are below the term energies of the observed levels. A comparison of experiment and calculation is shown in Graph~(a) of Figure~\ref{figResidualHomo}. Here, the differences of energy spacings
\begin{equation}
\begin{array}{ll}
\left(\nu^i-\nu^0\right)-\left(T_e^i-T_e^0\right)& {\rm for}\ i=1,\dots,N
\label{eqDifferencesResiduals}
\end{array}
\end{equation}
are plotted over the term energy $T_e^i$. The residuals of the energy spacings are up to seven times the experimental uncertainty for relative frequency measurements and a trend is clearly visible. Due to the wide range of term energies for the fit in \cite{falke206}, this is not surprising. However, for the comparison with the heteronuclear dimer, an accurate reproduction of the energy spacing as determined in this work is desirable.

For a least-squares fit of the energy spacings we consider the full potential but we allow only adjustments of the inner-branch  ($R<3.025$~\AA, binding energies smaller than $300~{\rm cm}^{-1}$). This leads to a little altered phase of the wavefunction in the long range region and hence to a modified ladder of levels. At short internuclear distances, the repulsive wall is modeled by the form 
\begin{equation}
A_1+A_2 R^{-C}
\end{equation}
where $A_1$ and $A_2$ ensure a continuously differentiable connection to the potential well. By varying $C$ and impelled the parameters $A_1$ and $A_2$ the residuals are reduced as shown in Graph~(b) of Figure~\ref{figResidualHomo}.

No more trends in the residuals are visible and the standard deviation is within $\pm 2~{\rm MHz}$, which is consistent with the experimental uncertainty of line positions considering the linearity of the laser scan and the reached signal-to-noise ratio in relation to the linewidth. The precision of the frequency scale is determined by the Fabry-P{\'e}rot interferometer used for interpolation. We verified that long term drifts of the free spectral range between the recording of scans and short term fluctuations during single scans are not limiting the precision. The energy position of all levels used in \cite{falke206} is still properly described.

\subsection{Scaling to the Heteronuclear Dimer}
\label{secFitHeteroAsymp}

\begin{figure}
\begin{flushleft}
\resizebox{1.0\columnwidth}{!}{\includegraphics{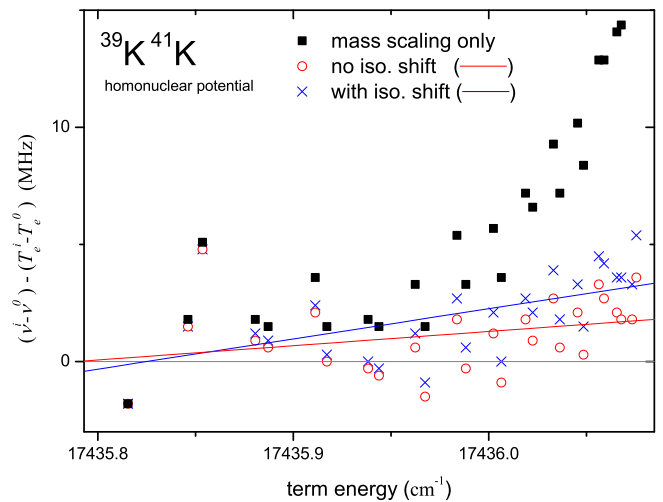}}
\end{flushleft}
\caption{Residuals of the observed and calculated energy spacing in the asymptotic region for three models. The potential curve is taken from the analysis of $^{39}{\rm K}_2$ and three different asymptotic behaviors are connected. The lines indicate remaining trends in the residuals.}
\label{figHeteroAsympFits}
\end{figure}

The adjusted potential energy curve is now applied to calculate term energies $T^i_e$ of the heteronuclear dimer. Three models are used to derive hyperfine potential energy curves that correspond to the ladders of asymptotic energies depicted in Figure~\ref{figAsympEnergies} and described above. The first model, which takes the hyperfine structure of $^{39}{\rm K}$ for both atoms (mass scaling only, left column in Figure~\ref{figAsympEnergies}), is clearly not able to reproduce the experimental findings (see squares in Figure~\ref{figHeteroAsympFits}). The other two models describe the observed energy spacing already within few MHz. Both take the smaller hyperfine structure of $^{41}{\rm K}$ into account, but only the third model (crosses in Figure~\ref{figHeteroAsympFits}) incorporates the atomic isotope shift. However, both models lead to residuals of the energy spacings that are systematically above zero. The deviations between model and observation become larger for spacings to higher vibrational levels (see lines in Figure~\ref{figHeteroAsympFits} that indicate linear fits of the residuals of these two models). However, the small differences already show that for scaling from one isotopomer to another the Born-Oppenheimer potentials reproduce the energy spacings of asymptotic levels ($v_{\rm A}=227$ to $v_{\rm A}$ around $240$), which are in the order of 2~GHz, to within 4~MHz. The absolute term energies of the observed levels of the heteronuclear dimer can be derived from the homonuclear dimer within the uncertainties of the experiment ($100~{\rm MHz}$ for the absolute frequency measurement and $50~{\rm MHz}$ for term energies of the ground state \cite{amiot95}), which relates to a relative accuracy of $2\times10^{-7}$ of the term values. Therefore, the vibrational assignment is known without uncertainty for $^{39}{\rm K}^{\,41}{\rm K}$ due to the knowledge of the full potential \cite{falke206}.

\subsection{Corrections to the Born-Oppenheimer Potential}
The observed trend in the residuals of the energy spacing for the least bound levels is a signature of the fact that the potential of the heteronuclear dimer cannot be derived completely from the potential of the homonuclear dimer by considering the appropriate potential asymptote from hyperfine interaction and atomic isotope shift. Quantitatively, the deviation of $4~{\rm MHz}$ from the model of the Born-Oppenheimer approximation can be compared to the statistical uncertainty of the last seven measurements (above $17\,436.05~{\rm cm^{-1}}$, see Fig.~\ref{figHeteroAsympFits}) of $2~{\rm MHz}/\sqrt{7}=760~{\rm kHz}$, where $2~{\rm MHz}$ is the uncertainty of a single measurement. We need to correct the Born-Oppenheimer potential in the inner part. The analysis of the Born-Oppenheimer approximation by Watson \cite{watson80} shows that the next higher order correction allows to use still the model of potentials for describung the nuclear motion, but the potentials become isotope dependent.

Thus, in the third step, we analyze now how the correction may be applied. We have concentrated on asymptotic levels but in our analysis we use a full potential. We now perform the least-squares fit for the energy spacing of the heteronuclear molecule. This time, we adjust the depth of the molecular potential derived from the homonuclear case by shifting the potential minimum with respect to the dissociation asymptote. The hyperfine potential energy curves and the long range part (including $C_3$, $C_6$ and $C_8$) of the potential are kept fixed for $R\ge 18.5$~\AA, whereas the potential well is shifted by a constant offset for $R < 18.5$~\AA. A continuous differentiable connection is ensured by automatic adjustment of very small $C_{10}/R^{10}$ and $C_{12}/R^{12}$ terms for $R\ge 18.5$~\AA.

In this way we assume that the correction to the Born-Oppenheimer potential is mainly a change of the depth of the potential. Small changes of the potential depth lead to changes in the phase of the vibrational wavefunction for the asymptotic region. Our experiment is therefore mainly measuring the accumulated effect of alternations of the potential but not sensitive to the excact $R$-dependence of the correction. In asymptotic methods \cite{gribakin93}, a simple approximation of the inner potential is usually done by starting the integration of the Schr{\"o}dinger equation for the vibration outwards from a certain internuclear distance with a defined phase as a starting condition. Simple functional relations are used in fits to adjust the starting condition as function of binding energy of the asymptotic levels. They release the researcher from modelling the inner part of the potential well. In our work, the vibrational wavefunction accumulates an additional phase mainly due to the constant offset. The $C_{10}$ and $C_{12}$ terms ensuring a continuous connection add also phase but these contributions are much more short range than $C_3$ and $C_6$, thus the additional phase shift is constant for the observed asymptotic levels.

\begin{figure}
\resizebox{1.0\columnwidth}{!}{\includegraphics{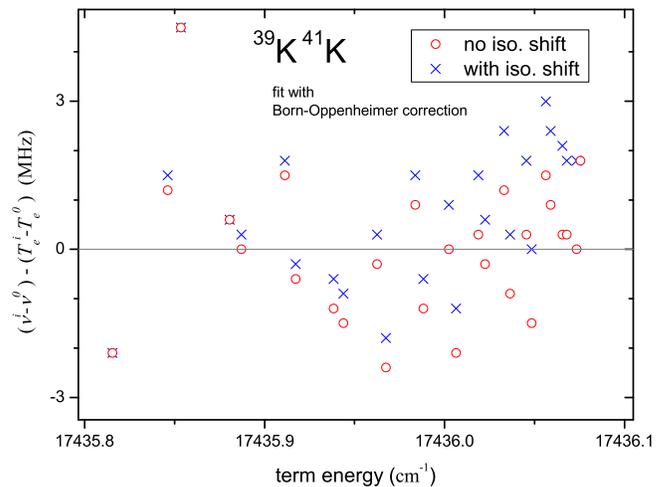}}
\caption{Residuals of a fit with Born-Oppenheimer corrections to the potential for $^{39}{\rm K}^{\,41}{\rm K}$. Two models were applied: with and without atomic isotope shift, see Figure~\ref{figAsympEnergies}.}
\label{figBOCorrection}
\end{figure}

For both asymptotic models (with and without isotope shift), a least-squares fit was performed for the asymptotic energy spacings of the heteronuclear dimer. The additional offset of the potential energy curve at small internuclear separations was the only free parameter. The obtained residuals are shown in Figure~\ref{figBOCorrection}.

Both fits result in residuals scattered around zero with a standard deviation of about $\pm 2$~MHz. No trends are visible in both cases any more. The applied change of the depth of the potential are comparative: deeper by around $200~{\rm MHz}$. The effective change of the level position in the observed interval is however only few ${\rm MHz}$. Both models lead to similar standard deviations of the fits. Therefore, we could not find a clear signature of the influence of the isotope shift, which would lead to a change-over from a resonant to detuned dipole-dipole character.

The change-over in the asymptotic behavior at the first excited electronic asymptote is hidden for the heteronuclear $^{39}{\rm K}^{\,41}{\rm K}$ through the hyperfine structure. The adiabatic potential energy curves for the different hyperfine states of the A~state are all connected to the two lowest asymptotic energies, which are indicated by circled numbers in Figure~\ref{figAsympEnergies}. The potential energy curves end for the two models almost at the same energy since the smaller hyperfine structure of $^{41}{\rm K}$ is partly compensated by the isotope shift (see \textcircled{3}, \textcircled{4} and \textcircled{5}, \textcircled{6}).

On the other hand, we show that in addition to a change of the reduced mass and the atomic hyperfine structure a correction to the potential energy curve is needed if one wants to derive the asymptotic level structure of one isotopomer from another with a precision better than $10~{\rm MHz}$ for the discussed isotope combinations.

\section{Conclusion}
\label{conclusion}
In a high resolution laser-spectroscopic experiment with a molecular beam, we investigated asymptotic levels of the A~$^1\Sigma_u^+$ state of the two isotopomers $^{39}{\rm K}_2$ and $^{39}{\rm K}^{\,41}{\rm K}$ up to the dissociation limit. An analysis of the level spacings revealed that for a correct asymptotic description the hyperfine structure of the isotopes has to be included properly. Applying as a first step of Born-Oppenheimer correction the atomic isotope shift between $^{39}{\rm K}$ and $^{41}{\rm K}$ of the $D$~lines is not sufficient. We observed that the inner Born-Oppenheimer potential derived from one isotopomer needs to be corrected to reproduce the observations of the second isotopomer within the experimental uncertainty of $2~{\rm MHz}$.

For the analysis, two models of the asymptotic energies were applied. One of them incorporates the atomic isotope shift that induces a changeover of a $R^{-3}$ behavior of a fine and hyperfine free long range potential to a $R^{-6}$ behavior. A $R^{-3}$ behavior is characteristic for homonuclear dimers whereas a $R^{-6}$ behavior is expected for heteronuclear dimer for binding energies up to the order of the isotope shift. The two models (hyperfine structure of $^{39}{\rm K}$ and $^{41}{\rm K}$ only or adding atomic isotope shift) lead to descriptions of our observations of similar quality. Therefore, we could not identify a signature of the expected changeover and we conclude that this effect is hidden by the hyperfine structure and the spin-orbit coupling since the atomic isotope shift is only $235~{\rm MHz}$.

The potential of the A~$^1\Sigma_u^+$ state is about $6300~{\rm cm}^{-1}$ deep and we applied a change of the potential depth of around $200~{\rm MHz}$ ($0.007~{\rm cm}^{-1}$). The Born-Oppenheimer potential is corrected by a relative amount in the order of $10^{-6}$, which is in the same order as the change of the reduced mass of the system of one electron interacting with the core of a potassium dimer, either $^{39}{\rm K}_2$ or $^{39}{\rm K}^{\,41}{\rm K}$ thus has the same magnitude as the normal mass shift in atomic physics. This correction of the Born-Oppenheimer potential can lead to significant changes of properties describing cold collisions, e.g., scattering lengths or positions of Feshbach resonances: a change of an asymptotic Feshbach resonance by $1.5~{\rm MHz}$ corresponds to a change of $1~{\rm Gauss}$ for a difference of atomic and molecular magnetic moments by one Bohr magneton. The resulting effect on a scattering length does not change linearly with the asymptotic energy levels. For large scattering lengths the binding energy of the last level is typically below $1~{\rm MHz}$. Thus, changes in such order can \emph{change the sign of the scattering length}. Only in cases of small scattering lengths the influence by Born-Oppenheimer corrections can be expected to be also small. The scattering length can be regarded as small if it is comparable to the characteristic length $(2\mu C_n/\hbar^2)^{1/(n-2)}$ \cite{tiesinga96}.

In our experimental setup, we will study the ground state asymptote of the two isotopomers $^{39}{\rm K}_2$ and $^{39}{\rm K}^{\,41}{\rm K}$ in order to describe cold collisions of the two combinations of isotopes and to study to which precision properties like scattering lengths or Feshbach resonance positions may be scaled by mass and hyperfine energies without a correction to the Born-Oppenheimer potential.

Such experiments can be done by adding a third laser to drive the decay \textcircled{4} in Figure~\ref{figSetup} coherently to the continuum and to scattering resonances. In a long term, we plan similar investigations on dimers of alkaline earth metals, which do not have hyperfine structure due to the lack of a nuclear moment and thus will show the Born-Oppenheimer correction more directly

\section*{Acknowledgments}
We thank the Deutsche Forschungsgemeinschaft for supporting this work within the SFB~407 and the European Commission in the frame of the Cold Molecules TMR Network under contract No. HPRN-CT-2002-00290.


\end{document}